\documentclass[superscriptaddress,twocolumn,preprintnumbers,amsmath,amssymb,aps]{revtex4}

\pdfoutput=1
\usepackage[utf8]{inputenc}
\usepackage[USenglish]{babel}

\usepackage{amsmath,amssymb,amsthm,amsfonts}
\usepackage{MnSymbol}
\usepackage{mathrsfs}
\usepackage{slashed}
\usepackage{graphicx}
\usepackage{subfigure}
\usepackage{color,rotating}
\usepackage{dsfont}
\usepackage{setspace}
\usepackage{verbatim}
\usepackage{hyperref,doi}
\usepackage[export]{adjustbox}
\usepackage[normalem]{ulem}
\usepackage{bbold}

\usepackage[sort&compress]{natbib}

\makeatletter
\def\l@subsubsection#1#2{}
\makeatother

\newcommand{\beq}{\begin{equation}}
\newcommand{\eeq}{\end{equation}}
\newcommand{\beqa}{\begin{eqnarray}}
\newcommand{\eeqa}{\end{eqnarray}}
\newcommand{\bfc}{\begin{figure}[t]\begin{center}}
\newcommand{\efc}{\end{center}\end{figure}}

\def\Fig#1{fig.~\ref{#1}}
\def\fig#1{fig.~\ref{#1}}
\def\Eq#1{eq.~(\ref{#1})}
\def\eq#1{(\ref{#1})}

\def\0#1#2{\frac{#1}{#2}}  

\graphicspath{{./figures/}}

\newcommand{\be}{\begin{eqnarray}}
\newcommand{\ee}{\end{eqnarray}}

\newcommand*\chem[1]{\ensuremath{\mathrm{#1}}}

\begin{document}

\title{Exploring the Hubbard Model on the Square Lattice at Zero Temperature\\with a Bosonized Functional Renormalization Approach}

\author{Sebastian J. Wetzel} \affiliation{Institut f\"ur Theoretische
  Physik, Universit\"at Heidelberg, Philosophenweg 16, 69120
  Heidelberg, Germany}

\begin{abstract}
%
%
%
%
%
We employ the functional renormalization group to investigate the phase diagram of the $t-t'$ Hubbard model on the square lattice with finite chemical potential $\mu$ at zero temperature. A unified scheme to derive flow equations in the symmetric and symmetry broken regimes allows a consistent continuation of the renormalization flow in the symmetry broken regimes. At the transition from the symmetric regime to the symmetry broken regimes, our calculation reveals leading instabilities in the d-wave superconducting and antiferromagnetic channels. Furthermore, we find a first order transition between commensurate and incommensurate antiferromagnetism. In the symmetry broken regimes our flow equations are able to renormalize around a changing Fermi surface geometry. We find a coexistence of d-wave superconductivity and antiferromagnetism at intermediate momentum scales $k$. However, there is a mutual tendency of superconductivity and antiferromagnetism to repel each other at even smaller scales $k$, which leads to the eradication of the coexistence phase in the limit of macroscopic scales.
\end{abstract}

\maketitle



\section{Introduction}
The $t-t'$- Hubbard model \cite{Hubbard1963,Gutzwiller1963,Kanamori1963} is a promising model to describe the phase diagram of electrons in $\chem{CuO}$ planes as they occur in cuprate high-temperature superconductors. The most distinct phases are antiferromagnetism and d-wave superconductivity in close vicinity to each other\cite{Anderson1987,Miyake1986,Scalapino1986,Bickers1987,Lee1987,Millis1990,Monthoux1991,Scalapino1995,Bickers1989,Bulut1993,Maier2000,Grote2002,Hankevych2002,Senechal2005,Maier2006,Maier2007,Scalapino2012}.
 Early works could already infer d-wave superconductivity by scaling arguments \cite{Schulz1987,Dzialoshinskii1987,Lederer1987}. More advanced purely fermionic renormalization calculations revealed the leading instabilities of the Hubbard model in the antiferromagnetic and in the d-wave channel \cite{Zanchi1996,Zanchi1998,Halboth2000,Halboth2000a,Salmhofer2001,Honerkamp2001,Honerkamp2001a,Katanin2005,Metzner2012,Husemann2009,Husemann2012,Uebelacker2012,Eberlein2015,Giering2012,Lichtenstein2017,Veschgini2017a}. While these methods have problems to calculate results in the symmetry broken phases, a combination of different schemes like renormalization group and mean field calculations can give insights in the interplay of different orders \cite{Metzner2006,Reiss2007,Eberlein2014,Wang2014,Yamase2016,Veschgini2017}. Our approach builds upon a bosonized renormalization group analysis of the Hubbard model which allows the renormalization flow to enter symmetry broken phases \cite{Baier2000,Baier2004,Baier2005,Krahl2007,Krahl2009,Krahl2009a,Friederich2010,Friederich2011}. For this purpose we employ a scale dependent Hubbard-Stratonovich transformation \cite{Hubbard1959,Stratonovich1957} which is generated during the renormalization flow \cite{Gies2002,Floerchinger2009}.
 
We employ a bosonized functional renormalization group calculation to examine the phase diagram of the $t-t'$ Hubbard model with finite chemical potential $\mu$ at zero temperature. We formulate a consistent set of flow equations in the symmetric and symmetry broken phases, which are supplied by a renormalization scheme which works in the symmetric and all symmetry broken phases. This allows us to renormalize around a changing Fermi surface in the magnetic symmetry broken phase.

The initial conditions of our flow equations are given by the fermionic Hubbard action in the Matsubara formalism, \Eq{eq:HubbardAction}. By examining the leading instabilities of the flow equations in the symmetric regime, we identify three different channels in which symmetry breaking occurs, see \Fig{fig:PhaseDiagOrd3}. These correspond to commensurate antiferromagnetism, incommensurate antiferromagnetism and d-wave superconductivity. An advantage of our method is the way of identifying incommensurate antiferromagnetism. This is because the the minimum of the inverse magnetic propagator is directly related to the type of antiferromagnetism. This property allows for an accurate examination of the transition between commensurate and incommensurate antiferromagnetism, see \Fig{fig:ICTransition}, which we identify as a transition of first order.
By continuing the flow equations into the regimes of symmetry breaking, we are able to examine the interplay of the magnetic and d-wave order parameters. We find that at intermediate scales $k$ there is a regime of coexisting antiferromagnetic and d-wave condensate. Our results suggest that these two phases have a tendency to repel each other, which leads to the vanishing of the coexistence phase at macroscopic scales, see \Fig{fig:PhaseDiagBroken}.


\section{Field Theoretic Formulation of the Hubbard Model}

\subsection{Hubbard Model}
The $t-t'$ Hubbard model on the square lattice is defined by the Hamiltonian
\begin{align}
H=-\sum_{ij,\sigma}t_{ij}c_{i,\sigma}^\dagger c_{j,\sigma}^{\phantom{\dagger}}+h.c. +U\sum_i (c_{i,\downarrow}^\dagger c_{i,\downarrow}^{\phantom{\dagger}})(c_{i,\uparrow}^\dagger c_{i,\uparrow}^{\phantom{\dagger}}) \ , 
\end{align}
where $t_{ij}=t$ for nearest neighbors, $t_{ij}=t'$ for next-to-nearest neighbors and $t_{ij}=0$ otherwise. We define the energy scales by setting $t=1$. The fermionic dispersion relation in momentum space is 
\begin{align}
\xi(Q)=-\mu-2t(\cos(q_x)+\cos(q_y))-4t' (\cos(q_x)\cos(q_y)) \ ,
\end{align}
where we have already included the chemical potential $\mu$, which denotes the level of doping.

\subsection{Functional Renormalization Group}
We address the phase diagram of the Hubbard model by calculating the quantum effective action $\Gamma$ via the flow equation for the effective average action $\Gamma_k$ \cite{Wetterich1993}
\begin{align}
\partial_k \Gamma_k= \frac{1}{2}\text{STr}\left( \Gamma^{(2)}_k+R_k\right)^{-1}\partial_k R_k = \frac{1}{2}\text{STr}\,\tilde\partial_k\ln\left( \Gamma^{(2)}_k+R_k\right) \, .
\label{eq:wettericheq}
\end{align}
We also introduce a short form notation of the scale derivative $\tilde \partial_k= (\partial_k R_k)\partial_{R_k}$ acting only on the regulator. For small $k\rightarrow0$, the effective average action becomes the full effective action $\Gamma$. In the limiting case of large scales $k\rightarrow \Lambda$, the effective action equals the microscopic action $\Gamma_{k}\rightarrow S$. Thus the initial condition for our renormalization group calculation is defined by the Hubbard model action in the Matsubara formalism
\begin{align}
S=&\sum_Q \psi^\dagger(Q) \left( i \omega_Q+\xi(Q)\right)\psi(Q)\nonumber \\
&+\frac{U}{2}\sum_{Q_1,\dots,Q_4}\left(\psi^\dagger(Q_1)\psi(Q_2)\right) \left( \psi^\dagger(Q_3)\psi(Q_4)\right)\nonumber \\
&\hspace{1cm} \times \, \delta(Q_1-Q_2+Q_3-Q_4) \ .
\label{eq:HubbardAction}
\end{align}
The fields depend on a collection of momenta and the Matsubara frequency $Q=(\omega_Q,\vec{q})=(\omega_Q,q_x,q_y)$. The sum is a short hand notation for an integration over all momenta and Matsubara frequencies $\sum_Q=\int_{-\infty}^{\infty} \frac{d\omega }{2\pi}\int_{[-\pi,\pi]^2}\frac{dq^2}{(2\pi)^2}$. The electrons are written as four component Grassmann valued fields $\psi(Q)=(\psi(Q)_\downarrow,\psi(Q)_\uparrow)^T$. While the Matsubara frequencies are discrete at nonzero temperature, they take on continuous values at zero temperature. This has a strong effect on the flow equations: The fermionic propagator is no longer gapped by the lowest Matsubara frequencies $\omega=\pm\pi T$, thus fermionic fluctuations contribute even for small energy scales $k$. This fact is also responsible for the increased effect of the shape of the Fermi surface on the renormalization flow. Hence, special care needs to be taken when dealing with contributions close to the Fermi surface. Furthermore, in the bosonic sector at finite temperature, only the Matsubara zero mode contributes to the flow equations, which induces dimensional reduction at low scales $k\ll\pi T$. In this case, the theory can be influenced only by spatial fluctuations. These two finite temperature properties cannot be exploited in the derivation of flow equations at zero temperature and thus yield an extra challenge for our calculations.

\subsection{Truncation}

While the Hubbard action, \Eq{eq:HubbardAction}, is the initial condition for the flow equation, other couplings are generated during the renormalization flow. The average effective action can be decomposed into contributions with respect to their fermionic and bosonic content

\begin{align}
\Gamma_k=&\Gamma_{F,k}+\Gamma_{FB,k}+\Gamma_{B,k} \ . 
\end{align}
The fermionic part $\Gamma_{F,k}$ contains the Hubbard action \eq{eq:HubbardAction} and contributions $\Gamma_F^m,\Gamma_F^d$ mimicking the magnetic and d-wave contributions of the Hubbard interaction $U$. They absorb the respective momentum dependence of $U$ arising during the renormalization flow.
\begin{align}
\Gamma_{F,k}=&\sum_Q \psi^\dagger(Q)P_F(Q)\psi(Q)\nonumber \\
&+\frac{U}{2}\sum_{Q_1,\dots,Q_4}\left(\psi^\dagger(Q_1)\psi(Q_2)\right) \left( \psi^\dagger(Q_3)\psi(Q_4)\right)\nonumber \\
&\hspace{1cm} \times \, \delta(Q_1-Q_2+Q_3-Q_4)+\Gamma_F^m+\Gamma_F^d
\end{align}
While we keep the Hubbard interaction $U$ fixed, we allow for a fermionic wave function renormalization $Z_F$ in the kinetic term 
$P_F=Z_F \left(i \omega_Q+\xi(Q)\right)$. The fermionic momentum channels 
\begin{align}
\Gamma_F^m=&-\frac{1}{2}\sum_{Q_1,\dots,Q_4}\lambda_F^m(Q_1-Q_2)\delta(Q_1-Q_2+Q_3-Q_4)\nonumber  \\
&\hspace{0.5cm}\times \, \left( \psi^\dagger(Q_1) \vec{\sigma}\psi(Q_2)\right)\left( \psi^\dagger(Q_3) \vec{\sigma}\psi(Q_4)\right)
\end{align}
and 
\begin{align}
\Gamma_F^d=&-\frac{1}{2}\sum_{Q_1,\dots,Q_4}\lambda_F^d(Q_1+Q_3)\delta(Q_1-Q_2+Q_3-Q_4)\nonumber  \\
&\hspace{0.5cm}\times \, f_d((Q_1-Q_3)/2)f_d((Q_2-Q_4)/2)\nonumber  \\
&\hspace{0.5cm}\times \, \left( \psi^\dagger(Q_1) \epsilon\psi^*(Q_3)\right)\left( \psi^T(Q_2) \epsilon \psi(Q_4)\right)
\end{align}
will be kept zero during the renormalization flow. Their flow will be redefined as contributions to the Yukawa couplings $h_m,h_d$ of the magnetic and d-wave bosons by flowing bosonization, see appendix \ref{sec:flowbos}.

The purely bosonic part of the effective average action is defined by
\begin{align}
\Gamma_{B,k}=&\frac{1}{2}\sum_Q \vec{m}^T(-Q) P_m(Q)\vec{m}(Q)\nonumber\\
&+\sum_Q d^*(Q) P_d(Q)d(Q)\nonumber\\
&+\sum_X U_k(\rho_m,\rho_d) \ .
\end{align}
Here $\vec{m}$ describes a magnetic boson and $d$ a d-wave superconducting Cooper-pair. The effective potential $U$ depends on the symmetry invariants $\rho_m=\frac{1}{2}\vec{m}^T\vec{m}$ and $\rho_d=d^*d$. The kinetic contributions can be decomposed into a frequency dependent part and a momentum dependent part.
\begin{align}
P_m(Q)&=Z_m \omega_Q^2+A_m F_m(Q)\nonumber \\
P_d(Q)&=Z_d \omega_Q^2+A_d F_d(Q) \ .
\end{align}
The minimum of the spatial shape factor $F_d$ is found at $(0,0)$, however the minimum of the magnetic kinetic shape factor $F_m$ can take different values for different kinds of magnetism. A minimum at $(0,0)$ denotes ferromagnetism, while a minimum at $\vec{\pi}=(\pi,\pi)$ denotes antiferromagnetism. In our phase diagrams we also find incommensurate antiferromagnetism, where the minimum is fourfold degenerate on the axis at $(\pi,\pi\pm\delta_{ic})$ and $(\pi\pm\delta_{ic},\pi)$.

We parametrize the bosonic propagators with functions which allow for an accurate examination of the expansion around their minima. The magnetic propagator in the case of commensurate antiferromagnetism is given by
\begin{align}
F_{m,c}(Q)=\frac{D_m |[\vec{q}+\vec{\pi}]|^2}{D_m+ |[\vec{q}+\vec{\pi}]|^2} \ , 
\end{align}
where $[\vec{q}]=((q_x +\pi\mod 2\pi)-\pi,(q_y+\pi\mod 2\pi)-\pi)$ denotes the projection of momenta into the 1st Brillouin zone $[-\pi,\pi]^2$. In the case of incommensurate antiferromagnetism this parametrization is enhanced by
\begin{align}
F_{m,ic}(Q)=\frac{D_m \tilde F(\vec{q})}{D_m+ \tilde F(\vec{q})} \ , 
\end{align}
where
\begin{align}
 \tilde F(\vec{q})=\frac{1}{4 \delta_{ic}}\left( (|[\vec{q}+\vec{\pi}]|^2-\delta_{ic}^2)^2+4[q_x+\pi]^2[q_y+\pi]^2\right)
\end{align}
is employed to expand around the incommensurate minimum. The d-wave propagator is similarly parametrized by
\begin{align}
F_{d}(\vec{q})=\frac{D_d |[\vec{q}]|^2}{D_d+ |[\vec{q}]|^2} \ .
\end{align}
The interactions in the bosonic sector are contained in the effective potential
\begin{align}
U_k(\rho_m,\rho_d)&=\sum_{n=1}^{\text{ord}}u_n \ ,
\label{eq:effectivepotential}
\end{align}
to various orders in $\rho_m$ and $\rho_d$. The lowest order of the effective potential 
\begin{align}
u_1&=\lambda_{10} (\rho_m-\rho_{m0})+\lambda_{01}(\rho_d-\rho_{d0})
\label{eq:masses}
\end{align}
contains the mass terms $\lambda_{10}=m_m^2$ and $\lambda_{01}=m_d^2$. They are finite in the symmetric phases and zero in the symmetry broken phases. In the latter case we expand the effective potential around the minimum at $(\rho_{m0},\rho_{d0})$. The second order interactions 
\begin{align}
u_2&=\frac{\lambda_{20}}{2} (\rho_m-\rho_{m0})^2+\lambda_{11} (\rho_m-\rho_{m0})(\rho_d-\rho_{d0})\nonumber \\
&\hspace{1cm}+\frac{\lambda_{02}}{2}(\rho_d-\rho_{d0})^2
\end{align}
determine the curvature around the expansion point. As long as $\text{det}_{BB}=\lambda_{20}\lambda_{02}-\lambda_{11}^2 >0$, the expansion point is a true minimum. The third order contributions are  
\begin{align}
u_3&=\frac{\lambda_{30}}{6} (\rho_m-\rho_{m0})^3+\frac{\lambda_{21}}{2} (\rho_m-\rho_{m0})^2(\rho_d-\rho_{d0})\nonumber \\
&\hspace{0.5cm}+\frac{\lambda_{12}}{2} (\rho_m-\rho_{m0})(\rho_d-\rho_{d0})^2+\frac{\lambda_{03}}{6}(\rho_d-\rho_{d0})^3 \ .
\end{align}
The interactions between the bosonic and the fermionic sector are mediated by the Yukawa couplings
\begin{align}
&\Gamma_{FB,k}=\nonumber \\
&-\sum_{Q_1,Q_2,Q_3} h_m(Q_1) \vec{m}(Q_1)\left( \psi^\dagger(Q_2)\vec{\sigma}\psi(Q_3) \right)\nonumber \\
&\hspace{1cm} \times \, \delta(Q_1-Q_2+Q_3)\nonumber \\ 
&-\sum_{Q_1,Q_2,Q_3} \frac{1}{\sqrt{2}}h_d(Q_1)f_d((Q_2-Q_3)/2) \nonumber \\
&\left(d^*(Q_1)( \psi^T(Q_2)\epsilon\psi(Q_3) -d(Q_1)( \psi^\dagger(Q_2)\epsilon\psi^*(Q_3) \right)\nonumber \\
&\hspace{1cm}\times \, \delta(Q_1-Q_2+Q_3) \ .
\end{align}
We parametrize the magnetic Yukawa couplings by the momentum-weighted average of a ferromagnetic interaction and an antiferromagnetic interaction $h_m(Q)=\frac{|[\vec{q}]|}{\sqrt{2}\pi}h_m(\Pi)+\frac{\sqrt{2}\pi-|[\vec{q}]|}{\sqrt{2}\pi}h_m(0)$. While solving the flow equations at $T=0$, we find $\partial_kh_m(0)=0$ and thus  $h_m(0)=0$. Here $f_d(Q)=\frac{1}{2}(\cos(q_x)-\cos(q_y))$ is the d-wave form factor. The $1/\sqrt{2}$ prefactor of $h_d$ together with a redefinition of the d-wave boson into real fields $d=\frac{1}{\sqrt{2}}(d_1+id_2)$ is useful to treat the magnetic and d-wave bosons on an equal footing. Then they can be summarized in a common language in form of a $\text{O}(2)\times\text{O}(3)$ symmetric bosonic submodel.

The regulator function introduces an artificial mass to the kinetic terms of the bosonic and fermionic fields. In our regularization scheme it only acts on the spatial momentum dependent part
\begin{align}
A_m\, F_{m,k}(Q)&=A_m \, F_{m}(Q)+R_B(F_{m}(Q)) \nonumber \\
A_d\, F_{d,k}(Q)&=A_d\, F_{d}(Q)+R_B(F_{d}(Q)) \nonumber \\
Z_F\, \xi_{k}(Q)&=Z_F\, \xi(Q)+R_F(\xi(Q))
\end{align}
for slow momentum modes. In our work we chose the Litim regulator \cite{Litim2001}. It acts on the bosonic fields as
\begin{align}
&R_B(F_{m}(Q))=A_{m}\,(k^2-F_{m}(Q))\Theta(k^2-F_{m,d}(Q))\nonumber \\
&R_B(F_{d}(Q))=A_{d}\,(k^2-F_{d}(Q))\Theta(k^2-F_{d}(Q)) \ ,
\end{align}
and on the fermionic fields as
\begin{align}
R_F(\xi(Q))=Z_F\,\text{sign}(\xi(Q))(k-|\xi(Q)|)\Theta(k-|\xi(Q)|) \ .
\end{align}
The choice of the infrared cutoff function $R_k$ can be continuoulsy extended to the symmetry broken regimes which is discussed in section \ref{sec:fermisurface}. 

\section{Leading Instabilities in the Symmetric Regime}

We solve the set of flow equations, see appendix \ref{sec:floweqs}, in the symmetric and symmetry broken regimes from $\ln(\Lambda)=\ln(k)=12$ to $\ln(k)=-8$. The initial condition in the form of the Hubbard action \eq{eq:HubbardAction} dictates the initial conditions for all other couplings. In the fermionic sector the wave function renormalization starts at $Z_F=1$. The bosonic sector is initially completely decoupled from the fermions, hence $h_m(Q)=h_d=0$. The bosons start at their Gaussian fixed points, hence $\lambda_{10}=\lambda_{01}=1$, $P_m(Q)=P_d(Q)=0$ and $\lambda_{ij}=0$ if $i+j>1$. Exemplary we plot the flow of the most important quantities at $\mu=-0.4,t'=-0.1$ in \fig{fig:FlowPlot}. The parameters correspond to a region in the $\mu,t'$-diagram, \fig{fig:PhaseDiagOrd3}, where the renormalization group flow enters all four possible regimes: the symmetric regime, antiferromagnetism, d-wave superconductivity and coexistence regime of antiferromagnetism and d-wave superconductivity.

\subsection{Phase Diagram of Leading Instabilities}

\begin{center}
\begin{figure}[htb!]
\includegraphics[width=0.45\textwidth]{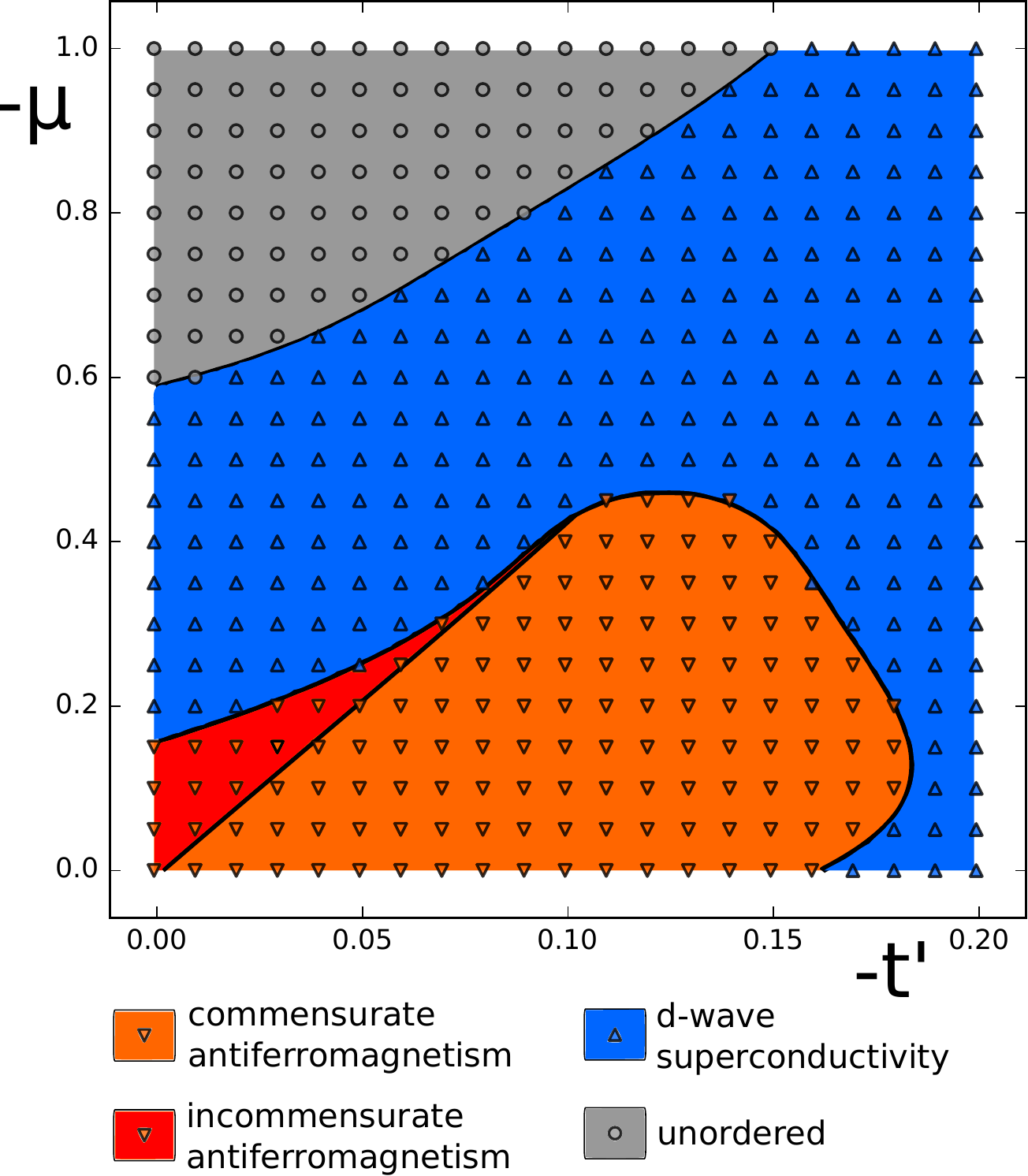}\\
\caption{Leading instabilities in the Hubbard model. The diagram was obtained by solving the renormalization flow equations in the symmetric phase for each $\mu -t'$. The truncation includes the effective potential expanded up to $\rho^3$.}
\label{fig:PhaseDiagOrd3}
\end{figure}
\end{center}

In this section we examine the solutions of the flow equations in the symmetric regime, in order to obtain a phase diagram of leading instabilities in the Hubbard model on the square lattice. More precisely, a leading antiferromagnetic instability occurs if $\lambda_{10}$ vanishes at a symmetry breaking scale $k_{SB}>0$, while $\lambda_{01}$ is still positive at $k_{SB}$. Similarly, a leading superconducting instability is characterized by $\lambda_{01}$ vanishing first. The boundary between antiferromagnetism and superconductivity is found where $\lambda_{10}$ and $\lambda_{01}$ vanish simultaneously at a common symmetry breaking scale $k_{SB}$. Finally, in the unordered phase $\lambda_{10}$ and $\lambda_{01}$ remain positive for $k\rightarrow0$. The diagram of leading instabilities corresponds to phase diagrams typically computed by purely fermionic flow equations. Indeed, integrating out the bosonic fields leads to a diverging four fermion interaction in the corresponding channels. In the vicinity of the transition between the antiferromagnetic and the superconducting regions the diagram of leading instabilities does not correspond to the true zero temperature phase diagram, since it does not capture the interplay between the two orders. The full phase diagram can only be obtained by following the flow in the spontaneously broken regime to $k\rightarrow 0$. This is discussed in section \ref{sec:broken}. The free parameters of our model are the chemical potenital $\mu$, the hopping parameters $t,\ t'$, and the Hubbard on-site interaction $U$. We set $t=1$ so that all other quantities are measured in units of $t$. The on-site interaction is set to $U=3$, which is lower than found in many cuprates. It is the choice for comparable renormalization group calculations, in order to avoid problems with too high interaction strengths. Previous experiments and calculations for cuprates agree on $U/t\approx6-8$ \cite{Coldea2001,Peres2002,Yang2017}. Even with lower interaction strength, the important mechanisms are already present and allow us to investigate the emergence of antiferromagnetism and superconductivity in the phase diagram of the Hubbard model from functional renormalization group calculations.

The diagram is calculated on a $21\times21$ grid containing values of $t'\in[0,-0.2]$ and $\mu\in[0,-1]$. The phase boundaries are statistically optimized using a machine learning algorithm called support vector machine. Doped cuprates, like $\chem{La_{2-x}Ba_xCuO_4}$ and $\chem{La_{2-x}Sr_xCuO_4}$, can be found for different levels of doping $x\rightarrow\mu$ at $t'/t\approx 0.14-0.17$ \cite{Pavarini2001,Delannoy2009}. Undoped, they exhibit antiferromagnetic order at low temperatures. However doping, or in our picture changing the chemical potential, leads to high-temperature superconductivity.

\begin{figure*}[htb!]
\includegraphics[width=0.9\textwidth]{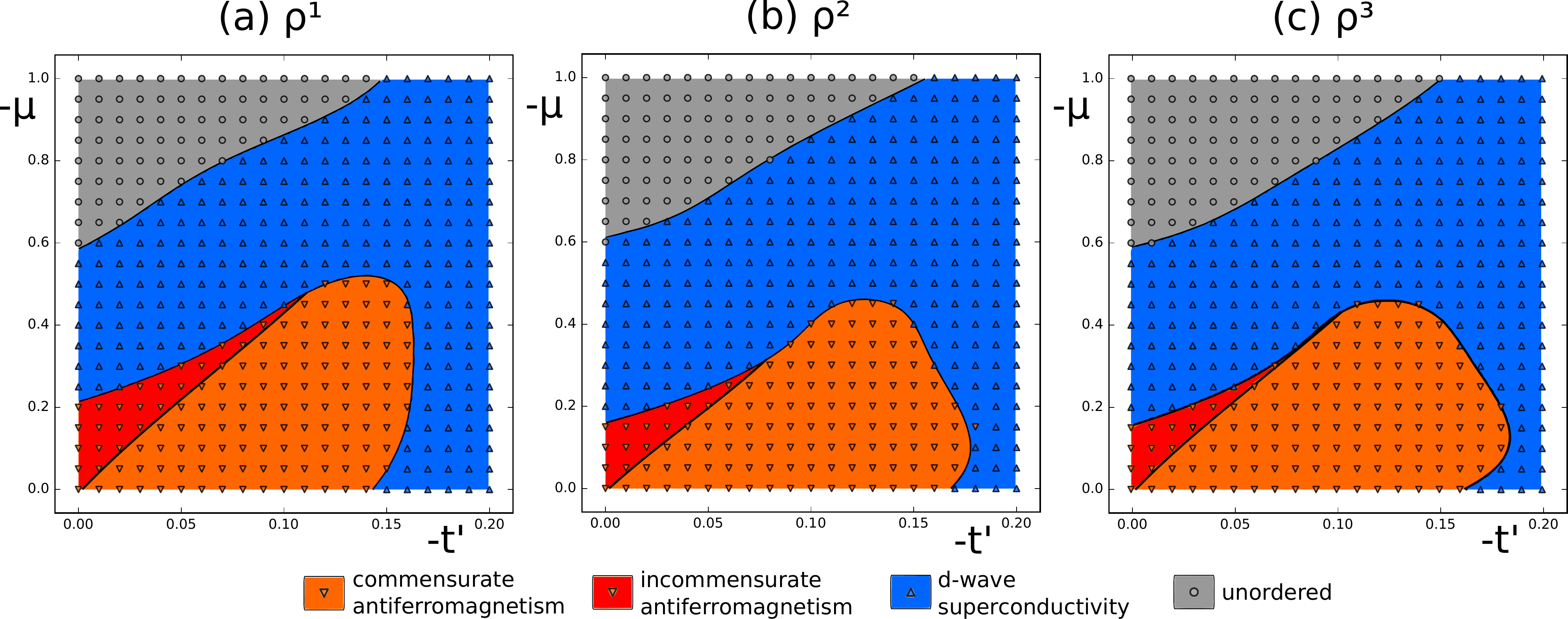}\\
\caption{Phase diagrams of leading instabilities calculated with different truncations of the effective potential: expansion up to (a) $\rho^1$,  (b) $\rho^2$,  (c) $\rho^3$.}
\label{fig:DiagConvergence}
\end{figure*}

The diagram of leading instabilities, \Fig{fig:PhaseDiagOrd3}, contains four different phases. At small chemical potential $\mu$ and small next-to-nearest neighbor hopping $t'$ there are two different antiferromagnetic regions. The commensurate region is separated from the incommensurate region by the line of Van-Hove singularities $\mu=4t'$. This transition is further analyzed in section \ref{sec:ICtransition}. At larger chemical potential and next-to-nearest neighbor hopping the leading instability is the d-wave superconductivity. In section \ref{sec:broken} we further demonstrate that it is possible to observe coexistence regions of both antiferromagnetism and d-wave superconductivity. Furthermore, at high chemical potential the ground state is unordered. We are aware that we might have underestimated the size of the antiferromagnetic region, where an accurate quantitative treatment requires the inclusion of charge density and s-wave bosons \cite{Friederich2011}. Along the Van-Hove line one always observes a broken symmetry in the antiferromagnetic or the d-wave channel. Our approach is not able to resolve a possible ferromagnetic instability \cite{Honerkamp2001b,Veschgini2017a} for large negative chemical potential and large negative next-to-nearest neighbor hopping on the Van-Hove line. This is a problem of momentum-shell schemes, like ours, and can be circumvented by temperature-flow renormalization techniques for the reasons explained in \cite{Honerkamp2001a}. Nevertheless, in all other properties the diagram of leading instabilities is in agreement with $\mu-t'-$diagrams from fermionic renormalization group calculations \cite{Veschgini2017}.

\subsection{Convergence of Diagrams}

It is a priori not clear to which order the effective potential, \Eq{eq:effectivepotential}, needs to be expanded to include all necessary effective interactions needed to calculate the phase diagram reliably. We calculated the diagram of leading instabilities for effective potentials to three different powers of the symmetry invariants $\rho_m, \rho_d$. In \Fig{fig:DiagConvergence} one can see on the left (a) a diagram containing only expansion terms to the order $\rho^1$, these are the bosonic mass terms, \Eq{eq:masses}. In the middle (b) the effective potential contains interactions up to $\rho^2$ or, in other words, up to quartic interactions for the fields. On the right (c) the diagram corresponds to the solution of the flow equation including terms up to order $\rho^3$. One can see a convergence of the phase diagrams for higher orders in $\rho$. We conclude that a truncation containing terms up to $\rho^2$ is sufficient to reliably calculate the phase diagram of the Hubbard model on the square lattice.

\begin{figure*}[htb!]
\includegraphics[width=0.9\textwidth]{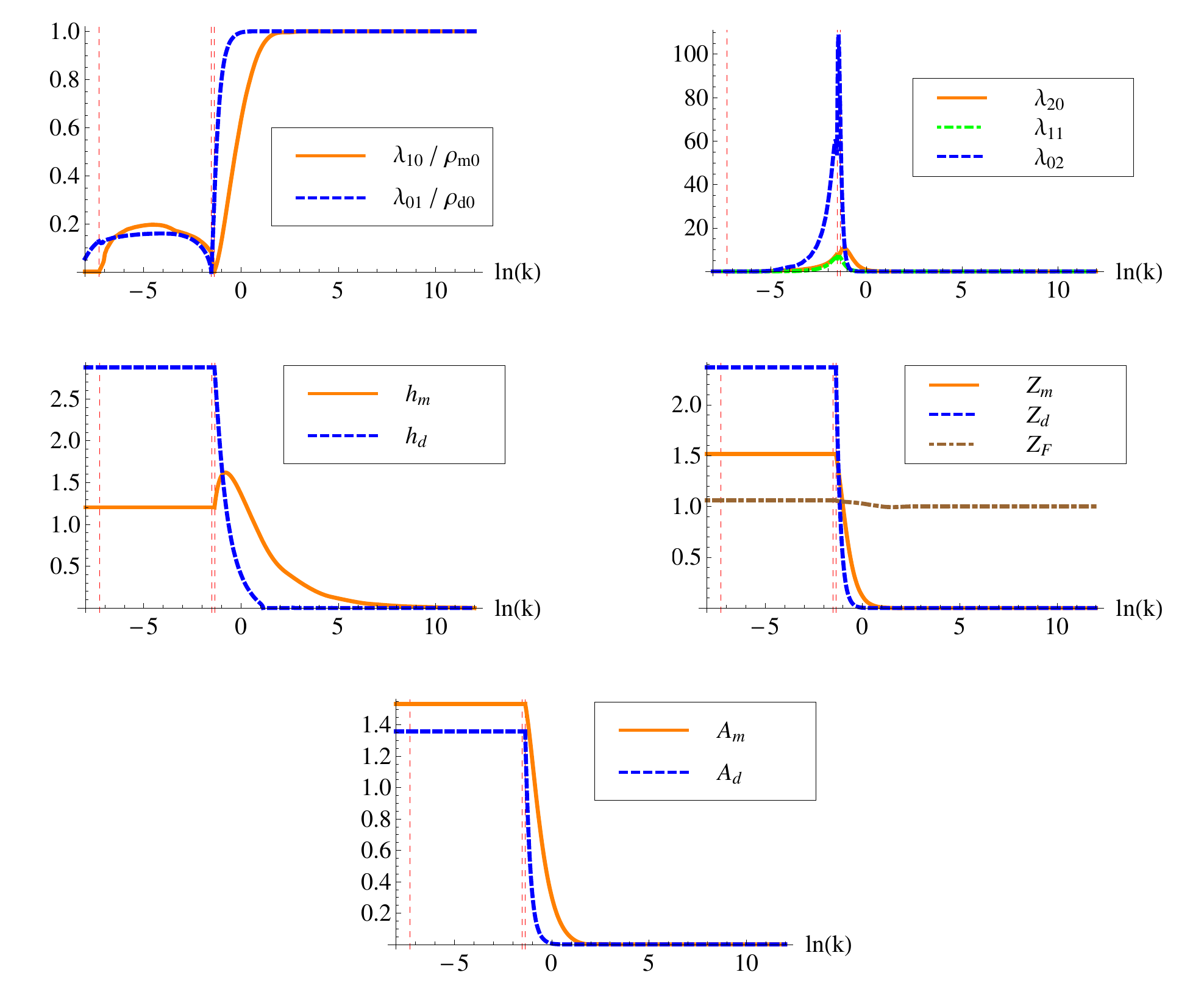}\\
\caption{Solutions of the flow equation at $\mu=-0.4,t'=-0.1$ from $\ln(\Lambda)=\ln(k)=12$ to $\ln(k)=-8$. The horizontal red dashed lines indicate transitions between symmetric and symmetry broken regimes. When the renormalization flow enters any symmetry broken phase, the flow of the masses $\lambda_{10},\lambda_{01}$ is continued by the flow of the minima of the effective potential $\rho_{m0},\rho_{d0}$. This particular flow trajectory visits all four possible regimes: the trajectory starts in the symmetric regime, then enters the antiferromagnetic regime, followed by a coexistence between d-wave and antiferromagnetism, afterwards the flow trajectory enters the d-wave regime. The minima of the effective potential $\rho_{m0}$ and $\rho_{d0}$ are rescaled $\times 20$.}
\label{fig:FlowPlot}
\end{figure*}

\subsection{Transition between Commensurate and Incommensurate Antiferromagnetism}
\label{sec:ICtransition}
\begin{figure*}[t!]
\includegraphics[width=0.9\textwidth]{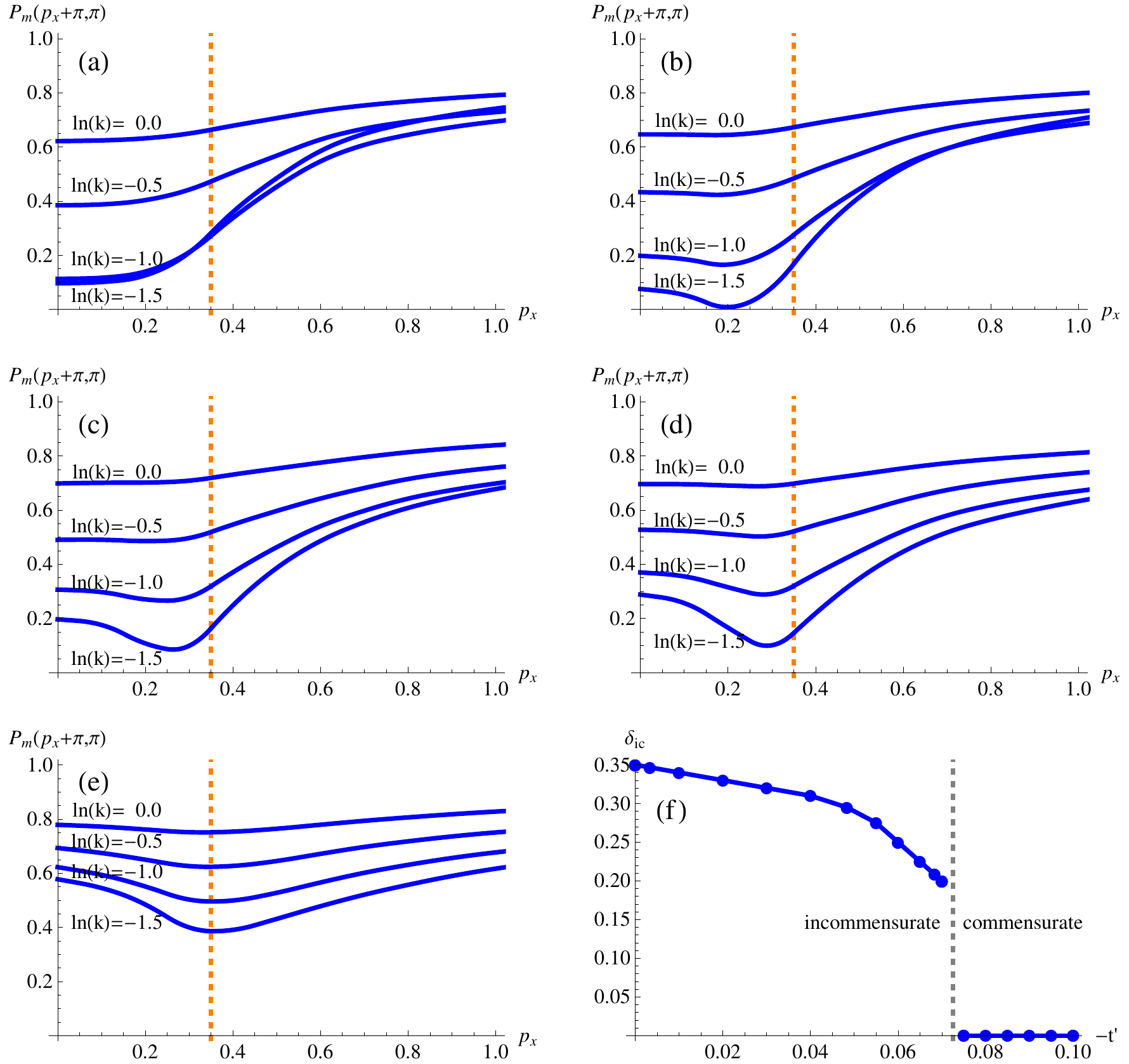}\\
\caption{First order transition between incommensurate and commensurate antiferromagnetism at fixed $\mu=0.3$. (a) $t'/t=-0.08$ (b) $t'/t=-0.07$ (c) $t'/t=-0.06$ (d) $t'/t=-0.05$ (e) $t'/t=-0.00$. The orange dotted line indicates the value of the incommensurate minimum at $t'=0$. (f) describes the transition of the minimum under a change in $t'$.}
\label{fig:ICTransition}
\end{figure*}

We examine the phase diagram of leading instabilities with a $16\times16$ resolution of the inverse magnetic propagator in the positive quadrant of the Brillouin zone $[0,\pi]^2$. We find that it can obtain two distinct minima. A minimum at $(\pi,\pi)$ corresponds to commensurate antiferromagnetism. A fourfold degenerate minimum at $(\pi,\pi+\delta_{ic})$ ,$(\pi,\pi-\delta_{ic})$ ,$(\pi+\delta_{ic},\pi)$ and $(\pi-\delta_{ic},\pi)$ corresponds to incommensurate antiferromagnetism. There is no other possible minimum, that the inverse magnetic propagator can obtain.

Earlier zero temperature calculations suggest that at vanishing next-to-nearest neighbor hopping $t'$ the incommensurability takes on the value $ \tilde\delta_{ic}(t'=0)=2\arcsin(|\frac{\mu}{2t}|)$\cite{Schulz1990}. Our results at the $t'=0$ line of the phase diagram are larger by $\approx 15\%$, $\delta_{ic}(t'=0)\approx2.3\arcsin(|\frac{\mu}{2t}|)$ . Considering for example $\mu=-0.3$, the result of the earlier work is $\tilde \delta_{ic}=0.301$, while our result is $\delta_{ic}\approx 0.35$, see \Fig{fig:ICTransition}. Our values of $\delta_{ic}(t'=0)$ are in agreement with earlier renormalization group studies \cite{Krahl2009a}.

The domains of commensurate antiferromagnetism and incommensurate antiferromagnetism are divided by a first order transition around the line $\mu\approx4t'$. This line corresponds to Van-Hove filling. In \Fig{fig:ICTransition} (a)--(e) we closely examine the transition for fixed $\mu=-0.3$ between the two regions, starting in the commensurate region by lowering $t'$. Each picture is calculated with 32 sampling points along the axis in the Brillouin zone. In the commensurate domain $\delta_{ic}(t'>0.075)=0$ (a) the magnetic propagator begins to flatten as it approaches the transition. At $t'=0.075$ the system undergoes a first order transition, where a new minimum emerges at a finite distance $\delta_{ic}(t'=0.075)\approx2/3 \times \delta_{ic}(t'=0)$ from the commensurate minimum. This minimum moves away (b),(c),(d). Finally, it converges slowly to $\delta_{ic}(t'=0)= 0.35$ (e). Also note that there is no significant change of the minimum of the propagator during the renormalization flow. This first order transition was also found in earlier calculations in \cite{Yamase2016}, by examining the jump in the incommensurability $\delta_{ic}$.

\section{Renormalization Flow in the Symmetry Broken Regimes}
\subsection{Deformation of the Fermi Surface}
\label{sec:fermisurface}

\begin{figure*}[t!]
\includegraphics[width=0.9\textwidth]{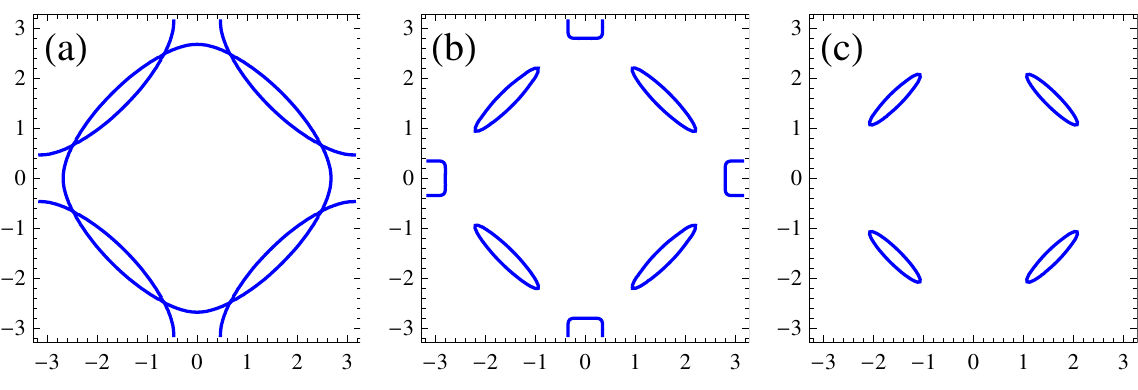}\\
\caption{The Fermi surface changes with increasing antiferromagnetic condensate. For next-to-nearest neighbor hopping $t'/t=-0.2$, chemical potential $\mu/t=-0.5$, fermionic wave function renormalization $Z_F=1$ and different sizes of the commensurate antiferromagnetic condensate (a) $\Delta_a=0$, (b) $\Delta_a=0.05$, (c) $\Delta_a=0.1$.}
\label{fig:FermiSurface}
\end{figure*}

\begin{center}
\begin{figure}[htb!]
\includegraphics[width=0.45\textwidth]{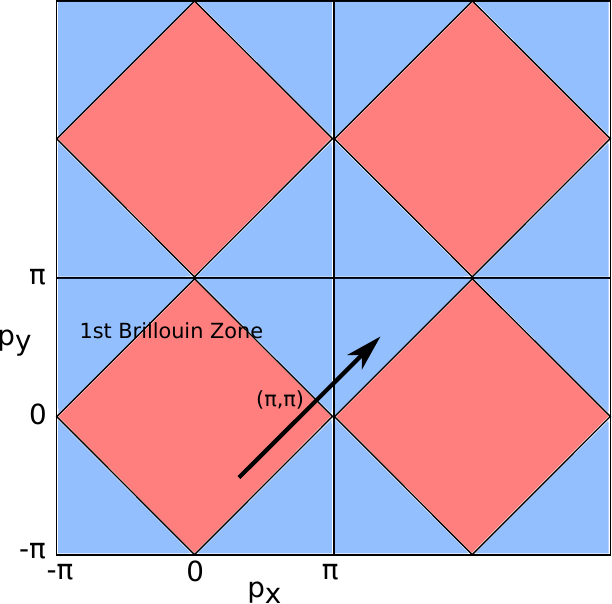}\\
\caption{The regulator \eq{eq:regulator} acts differently on the red and blue patches of the Brillouin zones. A shift by $\Pi=(\pi,\pi)$ induces a switch in the regulator function.}
\label{fig:Regulator}
\end{figure}
\end{center}

The Fermi surface in the antiferromagnetic broken phase changes its geometry according to the formula
\begin{align}
Z_F^2\xi(Q)\xi(Q+\Pi)-\Delta_a=0 \ , \Delta_a=2h_m(\Pi)^2\rho_a \ .
\label{eq:FermiSurface}
\end{align}
In \Fig{fig:FermiSurface} one can see how a finite antiferromagnetic condensate deforms the Fermi surface, such that Fermi pockets emerge. In the symmetric phase $\Delta_a=0$ the definition of the Fermi surface correctly reduces to $Z_F\xi(Q)=0$. As the fermionic propagator at zero temperature diverges at the Fermi surface, it naturally occurs in all fermionic contributions to the flow equations in the denominator. We now consider the effects of the deformation of the Fermi surface on the flow equations. At finite temperature when there is a finite lowest fermionic Matsubara mode, the flow equations are still gapped at sufficiently large scales $k$ in the magnetic symmetry broken phase. Practically this means that the flow equations in the symmetry broken regimes at finite temperature can be solved approximately by regularizing around the Fermi surface belonging to the symmetric regime $Z_F\xi(Q)=0$. This is scheme is formulated by choosing the regulator, here the Litim regulator \cite{Litim2001},
\begin{align}
R_F(\xi(Q))=Z_F\,\text{sign}(\xi(Q))(k-|\xi(Q)|)\Theta(k-|\xi(Q)|) \ .
\label{eq:litim}
\end{align}
A major problem arises at zero temperature. The lowest fermionic Matsubara frequency is zero and can thus no longer introduce a gap in the fermionic propagator. At zero temperature the flow equations would diverge at the Fermi surface at any scale when employing the just introduced Litim regularization scheme. Thus, it is imperative to regularize around the correct Fermi surface \eq{eq:FermiSurface}. In order to capture the deformation of the Fermi surface in the flow equations, we introduce a set of regulators acting differently on different patches in the Brillouin zone, \fig{fig:Regulator}, and thus differently on $\xi(Q)$ and $\xi(Q+\Pi)$,
\begin{align}
&R_F(\xi(Q))=\nonumber \\
&\hspace{0.1cm}\theta(\pi-|[q_x]|-|[q_y]|)\,Z_F\,\text{sign}(\xi(Q)-\frac{\Delta_a}{Z_F^2\xi_k(Q+\Pi)})\nonumber \\
&\hspace{0.1cm}\times(k-|\xi(Q)-\frac{\Delta_a}{Z_F^2\xi_k(Q+\Pi)}|)\Theta(k-|\xi(Q)-\frac{\Delta_a}{Z_F^2\xi_k(Q+\Pi)}|)\nonumber \\
&\hspace{0.1cm}+\theta(-\pi+|[q_x]|+|[q_y]|)Z_F\,\text{sign}(\xi(Q))\nonumber \\
&\hspace{1cm}\times(k-|\xi(Q)|)\Theta(k-|\xi(Q)|)  \ . 
\label{eq:regulator}
\end{align}
In appendix \ref{sec:regulator} we describe how this regulator ensures a gapped propagator in the vicinity of the Fermi surface. This is possible since in all flow equations the contributions from $\xi(Q)$ and $\xi(Q+\Pi)$ occur in pairs. In addition, this regulator ensures that one approaches the Fermi-surface with the correct sign.

While there are other regulators that are capable of capturing the Fermi surface, and in agreement with all requirements of regulators \cite{Wetterich1993} there are more conditions to be fulfilled in order to obtain reliable physical results: (i) In the symmetric case and the microscopic limit the regulator must reduce to the free case, i.e. the regulator \eq{eq:regulator} reduces to the Litim regulator \eq{eq:litim}. We experimented with different regulators, not fulfilling this condition, which completely changed the results. (ii) The regulator must be the same in all regimes. It was shown that switching the regulator while solving the flow equations has a strong impact on the results and introduces severe non-physical artifacts \cite{Pawlowski2017}. (iii) The Fermi surface needs to be approached with the correct sign. (iv) The regulator may not contain any divergences itself. Our regulator \eq{eq:regulator} complies with all of these conditions.

\subsection{Flow Diagram in the Symmetry Broken Regimes}
\label{sec:broken}

\begin{center}
\begin{figure}[htb!]
\includegraphics[width=0.45\textwidth]{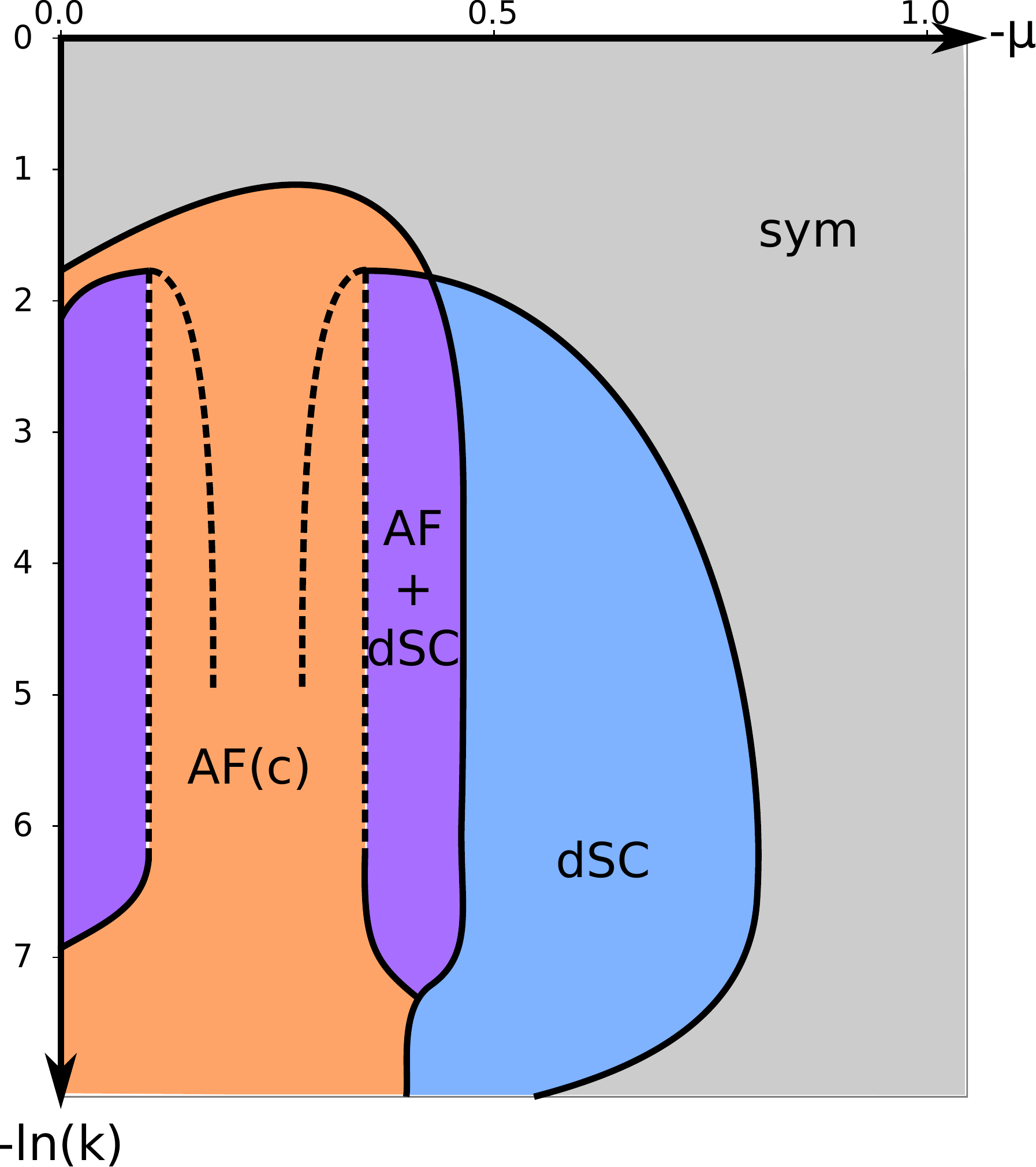}\\
\caption{Flow diagram at $t'=-0.1$. This diagram collects solutions for the flow equations for each $\mu$ and shows the regimes of finite condensates in dependence of the scale $k$. In the limit $k\rightarrow0$ or $\ln(k)\rightarrow-\infty$, the diagram shows the phases of the full effective action. Other scales $0<k<2\pi$ describe the physics in finite physical samples or local ordering. A coexistence phase is only present at intermediate scales.}
\label{fig:PhaseDiagBroken}
\end{figure}
\end{center}

In the symmetry broken phase we continued the flow of the effective potential including bosonic and fermionic fluctuations. The renormalization group flow is calculated within one consistent scheme of flow equations with which we also calculated the flow in the symmetric regime. The renormalization group flows of the wave function renormalizations $Z_m,A_m,Z_d,A_d$ and the flows of the Yukawa couplings $h_m,h_d$ are not continued. All flow equations are projected on commensurate antiferromagnetism in the symmetry broken phases. We solve the flow equations from $t=\ln(k)=12$ to $t=\ln(k)=-8$, where the numerical solution of the flow equation starts to break down.


In \Fig{fig:PhaseDiagBroken} we summarize the solutions of the flow equations for fixed next-to-nearest neighbor hopping $t'=-0.1$ for various values of the chemical potential $-\mu\in[0,1]$. This diagram corresponds to a collection of $\mu$-phase diagrams for different sizes of physical samples. This flow diagram contains, in addition to antiferromagnetism and d-wave superconductivity, regions of coexistence of these two phases. However, the coexistence vanishes in the limit of small momentum scales $k$ and thus large physical sample sizes. In other words, a coexistence phase in macroscopic physical systems can only be observed locally. The corresponding renormalization flow is shown in the flow of $\rho_{m0}$ and $\rho_{d0}$ of \Fig{fig:FlowPlot}. Here we see that both condensates emerge, however at scales $\ln(k)<-4$ they start to repel each other. Between the two coexistence regions in \Fig{fig:PhaseDiagBroken} there is a region where only antiferromagnetism is present. While in the middle at $t'\approx-0.025$ the d-wave mass $\lambda_{01}$ stays positive, there is a regime incapsulated by the dashed lines, in the vicinity of the coexistence phases, where this is not the case. Here the mass of the d-wave boson reaches a value of zero, however, the d-wave condensate is repelled immediately, such that the d-wave minimum $\rho_d$ of effective potential always stays zero. Even though our calculation is not quantitatively accurate in the symmetry broken phase, the mechanism responsible for the mutual repellence is independent of the exact sizes of both condensates. That is why we expect the eradication of the coexistence phase at macroscopic scales to be a rather robust result.

$\chem{La_2CuO_4}$ cuprates exhibit an orthorombic crystal structure with a very small asymmetry in lattice parameters in the $\chem{CuO_4}$ plane $a\approx b\approx5.4 \mathrm{\AA}$ and $c\approx 13.2 \mathrm{\AA}$ \cite{Longo1973}. It is possible to quantitatively relate the renormalization group scale $k$ approximately to physical scales. A scale of $k=2\pi$ corresponds to the size of a Brillouin zone and can thus be related to the physics of the size of one unit cell. Applying this deduction to \Fig{fig:PhaseDiagBroken} we find that superconductivity is most prominent at $\ln(k) \approx -7$ corresponding to a size of roughly $4\times10^4\mathrm{\AA}$. Furthermore, one can see that it is eradicated by allowing bosonic fluctuations at $\ln(k) < -9$ corresponding to $>3\times10^5\mathrm{\AA}$.

\subsection{On the Vanishing of Superconductivity}

From \Fig{fig:PhaseDiagBroken} one can infer that in the limit of $k\rightarrow0$ the superconductivity vanishes. The question arises if this observation bears any physical relevance or whether it is a pitfall of our limited truncation in the symmetry broken phases. We believe the true physical phase diagram shows superconductivity for large sample sizes. There are two possible scenarios to explain our results. First, our model could be incomplete in a sense that there would be the need for an effect limiting the scale of physical fluctuations to a size of roughly $10^5\mathrm{\AA}$ in order to still obtain a d-wave superconductivity at macroscopic scales. 

A second and  much more likely scenario is that our limited truncation in the symmetry broken phases limits the emergence of d-wave superconductivity. The d-wave condensate emerges quadratically from a growing Yukawa coupling $h_d$, see \Eq{eq:flowmd}. In the symmetry broken phases we do however not continue the flow of the Yukawa couplings. It is likely that $h_d$ grows larger as a natural continuation of the flow in the symmetric phase, see \Fig{fig:FlowPlot}. As a consequence, the d-wave minimum $\rho_{d0}$ would grow much larger than in our calculation and in turn could not be so easily destroyed by bosonic fluctuations.

While we cannot continue the renormalization flow of the d-wave Yukawa coupling, we examined if it would in principle be possible to obtain a non-vanishing d-wave condensate at even smaller scales $k$. We observe that the fermionic contributions, in the superconducting regime, always enhance the d-wave minimum $\rho_{d0}$, while only the bosonic fluctuations can reduce it. 

If we, for the moment, consider a sub-theory containing only bosonic fields and neglecting all fermionic contributions, this theory would, in the limit $k\rightarrow0$, become an effective three-dimensional statistical field theory. In accordance with the Mermin-Wagner theorem this theory can have a phase transition. Since fermionic contributions can only enhance the d-wave minimum $\rho_{d0}$, we conclude that the full theory containing bosonic and fermionic fields could in principle exhibit a phase transition. 

\begin{center}
\begin{figure}[htb!]
\includegraphics[width=0.45\textwidth]{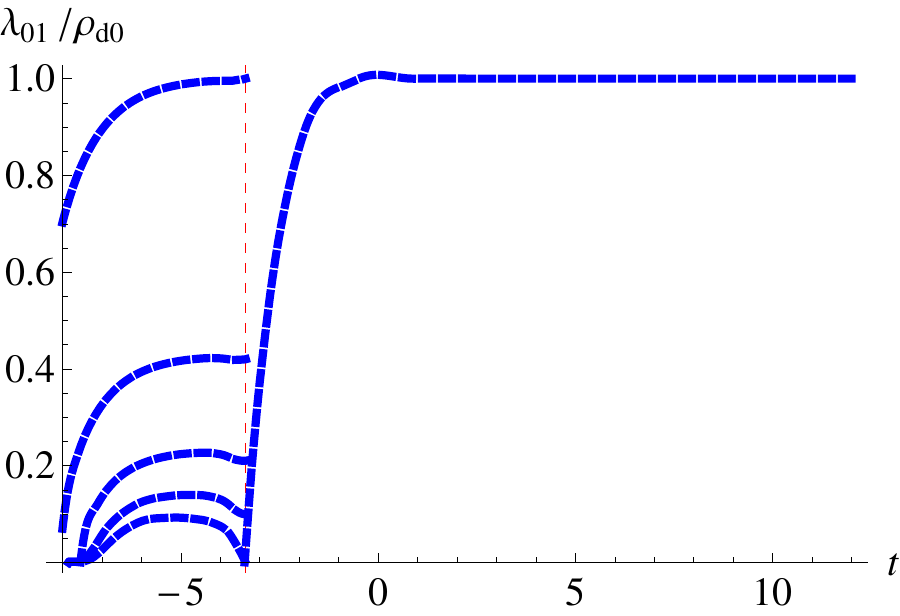}\\
\caption{Renormalization flow at $\mu=-0.66$, $t'=-0.1$ of the superconducting mass $\lambda_{01}$, which is continued with the flow of the superconducting minimum $\rho_{d0}$ below the symmetry breaking scale $k_{SB}$. The initial values of $\rho_{d0}$ are artificially enhanced at $k_{SB}$, from bottom to top: $\rho_{d0}=0$, $\rho_{d0}=0.001$, $\rho_{d0}=0.002$, $\rho_{d0}=0.004$, $\rho_{d0}=0.01$. The superconducting minimum of the effective potential $\rho_{d0}$ is rescaled $\times 100$.}
\label{fig:ManCond}
\end{figure}
\end{center}

The main effect of $h_d$ on the effective potential is the enhancement of the d-wave minimum of the effective potential $\rho_{d0}$. In \Fig{fig:ManCond} we compare the flow of the minimum $\rho_{d0}$ for different initial values at the symmetry breaking scale. One can see that in principle there exists a parameter range such that it is possible to be in the symmetry broken phase for an arbitrary finite scale $k$. However, the bosonic fluctuations tend to strongly reduce the d-wave condensate at any scale $\ln(k)<-7$.

\section{Conclusion}
In this article we explored the $\mu-t'-$phase diagram of the Hubbard model on the square lattice at zero temperature. A calculation in the symmetric phase revealed phases of commensurate antiferromagnetism, incommensurate antiferromagnetism and d-wave superconductivity, see \Fig{fig:PhaseDiagOrd3}. By comparing different truncations of the effective potential we examined the robustness of the results, see \Fig{fig:DiagConvergence}. Furthermore, our truncation allows for the continuation of the renormalization flow into the symmetry broken regimes, see \Fig{fig:FlowPlot}. We examined the transition between commensurate and incommensurate antiferromagnetism in detail. We find that it coincides with the Van-Hove line and is of first order, see \Fig{fig:ICTransition}. A nonzero antiferromagnetic condensate induces a continuous deformation of the Fermi surface, see \Fig{fig:FermiSurface}. Our regularization scheme \eq{eq:regulator} allows us to properly include fluctuations around the changing Fermi surface. Calculations in the symmetry broken regimes reveal a coexistence of antiferromagnetism and d-wave superconductity only on intermediate scales, see \Fig{fig:PhaseDiagBroken}. Beyond that, our results suggest that these phases have a tendency to repel each other. This mechanism leads to an eradication of the coexistence phase at macroscopic scales. A weakness of our calculations is the inability to decide the fate of the d-wave superconductivity at macroscopic scales, this can be resolved in the future by continuing the flow of the Yukawa couplings in the symmetry broken phases.

\emph{Acknowledgments} - We would like to thank Jan M. Pawlowski, Manfred Salmhofer, Michael M. Scherer, Kambis Veschgini and Christof Wetterich for useful discussions. We further thank Mathias Neidig and Shirin Nkongolo for proofreading the manuscript. S.W. acknowledges support by the Heidelberg Graduate School of Fundamental Physics.

\appendix
\section{Definitions}
\subsection{Pauli Matrices}

\begin{align}
\sigma_1=\begin{pmatrix}
0 &1\\
1&0
\end{pmatrix}\ \, \ 
\sigma_2=\begin{pmatrix}
0 &-i\\
i&0
\end{pmatrix}\ \, \ 
\sigma_3=\begin{pmatrix}
1 &0\\
0&-1
\end{pmatrix}
\end{align}

\subsection{Antisymmetric Matrix}
\begin{align}
\epsilon=\begin{pmatrix}
0 &1\\
-1&0
\end{pmatrix}\ \ , \ \epsilon^2=-\mathbb{1} 
\end{align}
\subsection{Dual Matrix}
\label{sec:dualmatrix}
If a matrix $A$ can be written in terms of Pauli Matrices $A=a \mathbb{1}+\vec{b}\vec{\sigma}$ and four real parameters $a,b_1,b_2,b_3$, its determinant evaluates to
\begin{align}
&\det(a \mathbb{1}+\vec{b}\vec{\sigma})=\begin{pmatrix}
a+b_3 &b_1-ib_2\\
b_1+ib_2&a-b_3
\end{pmatrix}=a^2-b_1^2-b_2^2-b_3^2\nonumber\\
&=\begin{pmatrix}
a-b_3 &-b_1+ib_2\\
-b_1-ib_2&a+b_3
\end{pmatrix}=\det(a \mathbb{1}-\vec{b}\vec{\sigma}) \ ,
\label{eq:dualmatrix}
\end{align}
such that we can define a dual matrix $A'=a \mathbb{1}-\vec{b}\vec{\sigma}$ with the same determinant. In the special case $\det A=1$, it follows $A \in SU(2)$.

\section{Flow Equations}
\label{sec:floweqs}

The derivation of the flow equation employs the flow equation of the effective average action, \Eq{eq:wettericheq}.

\subsection{Bosonic Contributions to the Effective Potential}

The flow equation of the effective potential can be decomposed into bosonic and fermionic contributions $\partial_kU=(\partial_kU)_F+(\partial_kU)_B$. The bosonic contributions correspond to the flow equations of a bosonic $\text{O}(2)\times\text{O}(3)$ model. They can be obtained in a straightforward manner by calculating the second field derivative of the effective action
\begin{align}
\Gamma_{BB}^{(2)}+R_{BB}=\begin{pmatrix}
R_m & 0 & 0 & I_{md} & 0 \\
0 & G_m &0 &0 & 0 \\
0&0 & G_m &0 & 0 \\
I_{md} & 0 &0 &R_d& 0\\
0 & 0 & 0 &0& G_d
\end{pmatrix} \ ,
\end{align}
which contains radial modes $R_m,R_d$, Goldstone modes $G_m,G_d$ and exchange modes $I_{md}$
\begin{align}
R_m&=P_m+2\rho_m U^{2,0}+U^{(1,0)}+R_B\nonumber \\
G_m&=P_m+U^{(1,0)}+R_B\nonumber \\
R_d&=P_d+2\rho_d U^{0,2}+U^{(0,1)}+R_B\nonumber \\
G_d&=P_d+U^{(0,1)}+R_B\nonumber \\
I_{md}&=2\sqrt{\rho_m \rho_d}U^{(1,1)}\nonumber \\
\text{det}_{BB}&=R_mR_d-I_{md}^2 \ .
\end{align}
Inverting this matrix
\begin{align}
(\Gamma_{BB}^{(2)}+R_{BB})^{-1}=\begin{pmatrix}
\frac{R_d}{\det_{BB}} & 0 & 0 & -\frac{I_{md}}{\det_{BB}} & 0 \\
0 & G_m^{-1} &0 &0 & 0 \\
0&0 & G_m^{-1}  &0 & 0 \\
 -\frac{I_{md}}{\det_{BB}} & 0 &0 &\frac{R_d}{\det_{BB}} & 0\\
0 & 0 & 0 &0&  G_d^{-1} 
\end{pmatrix} \ ,
\end{align}
immediately leads to the flow equation for the effective potential of an $\text{O}(2)\times\text{O}(3)$ model
\begin{align}
&(\partial_kU_k)_B=\nonumber \\
&\frac{1}{2}\sum_Q\left(\frac{R_m}{det_{BB}} +2 G_m^{-1}+\frac{R_d}{det_{BB}}+G_d^{-1} \right)\partial_kR_B \ .
\end{align}

\subsection{Fermionic Contributions to the Effective Potential}

The fermionic contribution to the effective potential 
\begin{align}
(\partial_k U)_{F}&=\frac{1}{2} \tilde \partial_k \text{STr} \ln \Gamma_{FF,k}^{(2)}(Q,Q')\nonumber \\
&=-\frac{1}{2} \tilde \partial_k \sum_{Q,Q'} \ln \det \Gamma_{FF,k}^{(2)}(Q,Q') \ ,
\end{align}
is much harder to obtain due to the momentum shift by $\Pi=(0,\pi,\pi)$ in the antiferromagnetic channel. The fermionic part of the second field derivative of the effective action 
\begin{align}
&(\Gamma_{FF}^{(2)}+R_{FF})(Q,Q')=\begin{pmatrix}
-h_d\epsilon (d_1-id_2) & h_m(\Pi) \vec{\sigma}\cdot \vec{m} \\
-h_m(\Pi)  \vec{\sigma}\cdot \vec{m} &h_d\epsilon  (d_1+id_2)
\end{pmatrix}\nonumber \\
&+\begin{pmatrix}
0 &-\mathbb{1}(P_F(-Q)+R_F(-Q))\\
\mathbb{1}(P_F(-Q)+R_F(-Q))&0
\end{pmatrix}
\end{align}
contains contributions from the fermionic propagator and the Yukawa couplings $h_m(\Pi),h_d$. It can be simplified by identifying similar terms
\begin{align}
&\Gamma_{FF,k}^{(2)}(Q,Q')=(\Gamma_{FF}^{(2)}+R_{FF})(Q,Q')=\nonumber \\
&\begin{pmatrix}
\tilde B(Q)\delta(Q-Q') &-A^T(-Q,Q')\\
A(Q,Q') &B(Q)\delta(Q-Q') 
\end{pmatrix}\ ,
\end{align}
where
\begin{align}
A(Q,Q')&=Z_F( i \omega_Q+\xi_k(Q))\delta(Q-Q')\nonumber \\
&\hspace{1cm}- \underbrace{h_m(\Pi)\, \vec{m}}_{\vec{M}}\cdot \vec{\sigma}\, \delta(\Pi-Q+Q')\nonumber \\
B(Q)&=h_d \,\epsilon\,  (d_1+id_2) \,f_d(Q)=D(Q)\epsilon\nonumber \\
\tilde B(Q)&=-h_d \,\epsilon \, (d_1-id_2) \,f_d(Q)=-\tilde D(Q)\epsilon \ .
\end{align}
A straightforward calculation that employs doubling the matrix in the argument of the determinant and employing its dual matrix \eq{eq:dualmatrix}
leads to
\begin{align}
&\det \Gamma_{FF,k}^{(2)}(Q,Q')\nonumber \\
&=\det \left( \Gamma_{FF,k}^{(2)}(Q,Q'')
\begin{pmatrix}
0&\mathbb{1}\\
\mathbb{1} &0
\end{pmatrix}
\Gamma_{FF,k}^{(2)}(-Q'',-Q')
\begin{pmatrix}
0&\mathbb{1}\\
\mathbb{1} &0
\end{pmatrix}
\right)^{1/2}\nonumber\\
&=\det \left( \tilde B (Q)B(-Q)\delta(Q-Q')+A(Q,Q'')A(-Q'',-Q')    \right)\nonumber\\
&=\det\Big( \underbrace{\left( \tilde D(Q) D(Q)+Z_F^2\omega^2+\xi_k(Q)^2+\vec{M}^2\right)}_{a(Q)}\delta(Q-Q') \nonumber\\
&\hspace{0.5cm}+\underbrace{(\xi_k(Q)+\xi_k(Q+\Pi))  \vec{M}   }_{\vec{b}(Q)}\cdot\vec{\sigma} \delta(Q-Q'+\Pi)\Big)\nonumber\\
&=\det \Big( \left(   a(Q)\delta(Q-Q'')-\vec{b}(Q)\cdot \vec{\sigma}\delta(Q-Q''+\Pi)                   \right)\nonumber\\
&\hspace{0.5cm} \times \left(   a(Q'')\delta(Q''-Q')+\vec{b}(Q'')\cdot \vec{\sigma}\delta(Q''-Q'+\Pi)\right) \Big)^{1/2} \nonumber\\
&=\det \left(   (a(Q)a(Q+\Pi)-\vec{b}(Q)\cdot\vec{b}(Q)     )\delta(Q-Q')            \right)^{1/2} \ ,
\end{align}
where we have used symmetry properties of the fermionic kinetic term $\xi_k(Q)=\xi_k(-Q)$ and the d-wave form factor $f_d(Q)=\frac{1}{2} (\cos(q_x)-\cos(q_y))=-f_d(Q+\Pi)$. Plugging all together and solving for $\omega$ yields
\begin{align}
&\det \Gamma_{FF,k}^{(2)}(Q,Q')=\det \big(J_+\,J_-\,\delta(Q-Q')\big)^{1/2} \ ,
\end{align}
where
\begin{align}
J_\pm&=Z_F^2 \omega^2 +\tilde D(Q)D(Q)+\Big( \frac{Z_F}{2}\big((\xi_k(Q)+\xi_k(Q+\Pi)\big)\nonumber \\
&\hspace{1cm}\pm \sqrt{\vec{M}^2+\frac{Z_F^2}{4}(\xi_k(Q)-\xi_k(Q+\Pi))^2}\ \Big)^2\nonumber \\
&=Z_F^2 \omega^2 +2 h_d^2 f_d(Q)\rho_d+\Big( \frac{Z_F}{2}\big((\xi_k(Q)+\xi_k(Q+\Pi)\big)\nonumber \\
&\hspace{1cm}\pm \sqrt{2 h_m(\Pi)^2 \rho_m+\frac{Z_F^2}{4}(\xi_k(Q)-\xi_k(Q+\Pi))^2}\ \Big)^2 \ .
\end{align}
The sum over $Q'$ can be performed trivially
\begin{align}
&\sum_{Q,Q'} \ln \det \Gamma_{FF,k}^{(2)}(Q,Q')=\sum_Q\frac{1}{2}(\ln J_+ + \ln J_-) \ ,
\end{align}
which concludes the calculation of the fermionic contribution to the flow equation of the effective potential
\begin{align}
(\partial_k U)_{F}&=-\frac{1}{2} \tilde \partial_k \sum_Q\frac{1}{2}(\ln J_+ + \ln J_-) \ .
\end{align}
\subsection{On Matsubara Integrals}

While the Matsubara integrals for higher order couplings can be integrated in a straightforward manner, the Matsubara integral for the effective potential itself evaluates to $\infty$. Since we are not interested in finite shifts of the overall energy minimum, we extract the finite part which depends on $\rho_m,\rho_d$ via
\begin{align}
&\int_{-\infty}^{\infty}d\omega \ln (Z_F^2 \omega^2 + A^2)\nonumber \\
&=\int_{-\infty}^{\infty}d\omega \underbrace{\ln (Z_F^2 \omega^2)}_{\infty}+\ln (1+ \frac{A^2}{Z_F^2\omega^2}) \ .
\end{align}
The finite contribution can then be safely integrated
\begin{align}
&=\int_{-\infty}^{\infty}d\omega \ln (1+ \frac{A^2}{Z_F^2\omega^2})\nonumber \\
&=\omega \ln (1+\frac{A^2}{Z_F^2 \omega^2})+\frac{2A}{Z_F}\arctan (\frac{Z_F \omega}{A})\big|_{-\infty}^{\infty}\nonumber \\
&=2 \pi \frac{A}{Z_F} \ .
\end{align}

\subsection{Flow Equations for the Bosonic Couplings}
We expand the effective potential \eq{eq:effectivepotential} around its minimum, thus we need to adjust the flow equations for a change in the minimum $\partial_k\rho_m,\partial_k\rho_d$.
\begin{align}
&\partial_k \lambda_{i,j}=\partial_{\rho_m}^{i}\partial_{\rho_d}^j\partial_k U(\rho_m,\rho_d)\big|_{\rho_{m0},\rho_{d0}}\nonumber \\
&=\left(\partial_{\rho_m}^{i}\partial_{\rho_d}^j \partial_k U(\rho_m,\rho_d)\right) \big|_{\rho_{m0},\rho_{d0}}\nonumber \\
&\hspace{0.5cm}+\left(\partial_{\rho_m}^{i+1}\partial_{\rho_d}^j  U(\rho_m,\rho_d)\right)|_{\rho_{m0},\rho_{d0}}\partial_k \rho_{m0}\nonumber \\
&\hspace{0.5cm}+\left(\partial_{\rho_m}^{i}\partial_{\rho_d}^{j+1}  U(\rho_m,\rho_d)\right)|_{\rho_{m0},\rho_{d0}}\partial_k \rho_{d0}
\end{align}
The flow equations of the minima of the effective potential are in the magnetic broken phase
\begin{align}
\partial_k\rho_{m0}&=-\frac{\partial_{\rho_m}(\partial_kU)}{\lambda_{20}}|_{\rho_{m0},\rho_{d0}} \ ,
\end{align}
in the d-wave broken phase
\begin{align}
\partial_k\rho_{d0}&=-\frac{\partial_{\rho_d}(\partial_kU)}{\lambda_{02}}|_{\rho_{m0},\rho_{d0}} \ ,
\end{align}
and in the regimes where both symmetries are broken
\begin{align}
\partial_k\rho_{m0}&=-\frac{\partial_{\rho_m}(\partial_kU)\lambda_{02}-\partial_{\rho_d}(\partial_kU)\lambda_{11}}{\lambda_{20}\lambda_{02}-\lambda_{11}^2}|_{\rho_{m0},\rho_{d0}}\nonumber \\
\partial_k\rho_{d0}&=-\frac{\partial_{\rho_d}(\partial_kU)\lambda_{20}-\partial_{\rho_m}(\partial_kU)\lambda_{11}}{\lambda_{20}\lambda_{02}-\lambda_{11}^2}|_{\rho_{m0},\rho_{d0}} \ .
\end{align}

\subsection{Flowing Bosonization}
\label{sec:flowbos}

Flowing bosonization induces a scale-dependent Hubbard-Stratonovich transformation \cite{Gies2002,Floerchinger2009}. In the present work we translate all diverging momentum channels of the Hubbard action U to Yukawa interactions mediated by magnetic and d-wave bosons. To this purpose we derive exemplary the flowing bosonization in magnetic channel. In $\Gamma_{m,k}^{\text{FHS}}$ we collect all couplings involved in the flowing bosonization.
\begin{align}
&\Gamma_{m,k}^{\text{FHS}}\nonumber \\
=& \frac{1}{2} \sum_Q \vec{m}^T(-Q)\left(P_m(Q)+m^2_m\right)\vec{m}(Q) \nonumber \\
&-\sum_{Q_1,Q_2,Q_3} h_m(Q) \vec{m}(Q_1) \cdot \left( \psi^\dagger(Q_2) \vec{\sigma}\psi(Q_3)\right)\nonumber  \\
&\hspace{0.5cm}\times \,  \delta(Q_1-Q_2+Q_3)\nonumber \\
&-\frac{1}{2}\sum_{Q_1,\dots,Q_4}\lambda_F^m(Q_1-Q_2)\delta(Q_1-Q_2+Q_3-Q_4)\nonumber  \\
&\hspace{0.5cm}\times \, \left( \psi^\dagger(Q_1) \vec{\sigma}\psi(Q_2)\right)\left( \psi^\dagger(Q_3) \vec{\sigma}\psi(Q_4)\right)
\end{align}
 We introduce an artificial scale-dependent bilinear field
\begin{align}
\tilde{\vec{m}}(P)_k= \sum_Q  \left( \psi^\dagger(Q) \vec{\sigma}\psi(P+Q)\right) \ ,
\end{align}
whose scale-dependence is chosen such that 
\begin{align}
\partial_k\vec{m}(Q)=\alpha_k^m(Q) \tilde{\vec{m}}(Q)  \ .
\end{align}
The function $\alpha_k^m(Q)$ will be specified later. Then we can rewrite $\Gamma_{m,k}^{\text{FHS}}$ in terms of the artificial bilinear
\begin{align}
&\Gamma_{m,k}^{\text{FHS}}\nonumber \\
=& \frac{1}{2} \sum_Q \vec{m}^T(-Q)\left(P_m(Q)+m^2_m\right)\vec{m}(Q) \nonumber \\
&- \sum_Q h_m(Q)\vec{m}^T(-Q)\tilde{\vec{m}}(Q) \nonumber \\
&-\frac{1}{2} \sum_Q \lambda_F^m(Q) \tilde{\vec{m}}^T(-Q)\tilde{\vec{m}}(Q) \ . 
\end{align}
The flow equation of the effective action obtains additional terms which arise from the scale-dependent bilinear
\begin{align}
\partial_k \Gamma_{k}=& \partial_k \Gamma_{k} |_{m_k}\nonumber \\
&+\sum_Q \alpha_k^m(Q) \vec{m}^T(-Q)\left(P_m(Q)+m^2_m\right)\tilde{\vec{m}}(Q)\nonumber\\
&-\sum_Q \alpha_k^m(Q) h_m(Q)\tilde{\vec{m}}^T(-Q)\tilde{\vec{m}}(Q) \ . 
\end{align}
The flow equations for the couplings $h_m$ and $\lambda_m^F$ obtain additional terms 
\begin{align}
\partial_k h_m(Q)&=\partial_k h_m(Q)|_{m_k}-\alpha_k^m(Q) \left(P_m(Q)+m^2_m\right)\nonumber \\
\partial_k \lambda_F^m(Q) &= \partial_k \lambda_F^m(Q)|_{m_k}+ 2 \alpha_m(Q) h_m(Q)\stackrel{!}{=}0 \ .
\end{align}
By choosing $\alpha^m_k(Q)$ such that the flow of $\lambda_m^F$ becomes zero, we induce an additional contribution to the flow of the Yukawa coupling $h_m$. A similar deduction can be done for the d-wave channel. Thus we arrive at modified flow equations 
\begin{align}
\partial_k h_m(Q)&=\partial_k h_m(Q)|_{m_k}+\frac{P_m(Q)+m_m^2}{2 h_m(Q)}\partial_k \lambda_F^m(Q)|_{m_k}\nonumber \\
\partial_k h_d(Q)&=\partial_k h_d(Q)|_{d_k}+\frac{P_d(Q)+m_d^2}{2 h_d(Q)}\partial_k \lambda_F^d(Q)|_{d_k} \ .
\label{eq:flowbos}
\end{align}
\subsection{Flow Equations for Yukawa Couplings}
The flow equations for $h_m$ and $h_d$ consist of a direct contribution and an indirect contribution arising from flowing bosonization, \Eq{eq:flowbos}. In the latter case the flow of $\lambda_F^m,\lambda_F^d$ is transformed into a contribution to the corresponding Yukawa coupling. When deriving the flow equations for $\lambda_F^m$ and $\lambda_F^d$, there arises an ambiguity in choosing the external momenta, we choose $L=(0,\pi,0)$ and $L'(0,0,\pi)$ in order to evaluate our contributions close to the Fermi surface. 
\begin{align}
\partial_k \lambda_F^m&=\frac{1}{3} \partial_k \Gamma^{(4)}_{F,ph}(L,L',-L,-L')\nonumber \\
\partial_k \lambda_F^d&=\frac{1}{2}\Big( \partial_k \Gamma^{(4)}_{F,pp}(L,L,-L,-L)\nonumber \\
& \hspace{1cm} -\partial_k \Gamma^{(4)}_{F,pp}(L,L',-L,-L') \Big)
\end{align}
As a real boson the magnetic channel collects all contributions arising from particle-particle diagrams. The d-wave boson describes Cooper pairs, which is why we collect the particle-particle diagrams in this channel. The prescription of how to extract the contributions to the d-wave coupling $\lambda_F^d$ was motivated in \cite{Krahl2009}.

\subsubsection{Magnetic Yukawa Coupling}
\begin{widetext}
\begin{align}
&\partial_k h_m(\Pi)^2  =  \tilde \partial_k\sum_Q \nonumber \\
& \frac{1}{6} \lambda_{10}\Big( \frac{8 h_d^2 F_{d,k}(L+Q/2)F_{d,k}(L'+Q/2) U (\omega^2-\xi_k(L+Q)\xi_k(L'+Q))}{Z_F^2(Z_d\omega^2+F_{d,k}(Q))(\omega^2+\xi_k(L+Q)^2)(\omega^2+\xi_k(L'+Q)^2)} \nonumber \\
&\hspace{1cm}+\frac{2 U^2 (\omega^2 -\xi_k(Q)\xi_k(Q+\Pi))}{Z_F^2(\omega^2+\xi_k(Q)^2)(\omega^2+\xi_k(Q+\Pi)^2)} \nonumber \\
&\hspace{1cm}+\frac{8h_d^4F_{d,k}(L+Q/2)F_{d,k}(L'+Q/2)F_{d,k}(\Pi/2+Q/2)(\omega^2-\xi_k(L+Q)\xi_k(L'+Q))}{Z_F^2(Z_d\omega^2+F_{d,k}(Q))(Z_d\omega^2F_{d,k}(\Pi+Q))(\omega^2+\xi_k(L+Q)^2)(\omega^2+\xi_k(L'+Q)^2)} \nonumber \\
&\hspace{1cm}-\frac{6h_m(Q+L)^2h_m(Q+L')^2(\omega^2-\xi_k(Q)\xi_k(Q+\Pi))}{Z_F^2(Z_m \omega^2 + F_{m,k}(L+Q))(Z_m \omega^2 + F_{m,k}(L'+Q))(\omega^2+\xi_k(Q)^2)(\omega^2+\xi_k(Q+\Pi)^2)}\Big) \nonumber \\
&+\frac{4 h_m(\Pi)^2h_d^2F_{d,k}(L+Q/2)F_{d,k}(L'+Q/2)(\omega^2-\xi_k(L+Q)\xi_k(L'+Q))}{Z_F^2(Z_d \omega^2+F_{d,k}(Q))(\omega^2+\xi_k(L+Q)^2)(\omega^2+\xi_k(L'+Q)^2)} \nonumber \\
&-\frac{2h_m(\Pi)^2h_m(Q)^2(\omega^2-\xi_k(L+Q)\xi_k(L'+Q))}{Z_F^2(Z_m\omega^2+F_{m,k}(Q))(\omega^2+\xi_k(L+Q)^2)(\omega^2+\xi_k(L'+Q)^2)}\nonumber \\
&+\frac{2h_m(\Pi) U (\omega^2-\xi_k(Q)\xi_k(Q+\Pi))}{Z_F^2(\omega^2+\xi_k(Q)^2)(\omega^2+\xi_k(Q+\Pi)^2)}
\end{align}
\end{widetext}

\subsubsection{D-Wave Yukawa Coupling}
\begin{widetext}
\begin{align}
&\partial_k h_d^2  = \tilde \partial_k \sum_Q \nonumber \\
& \frac{6 h_d^2F_{d,k}(Q) h_m(Q+L)^2}{Z_F^2(Z_m \omega^2+F_{m,k}(L+Q)(\omega^2+\xi_k(Q)^2))}\nonumber\\
&+\frac{1}{4}\lambda_{01}\frac{9h_m(Q+L)^4}{Z_F^2(Z_m\omega^2+F_{m,k}(L+Q))^2(\omega^2+\xi_k(Q)^2)}\nonumber\\
&-\frac{1}{4}\lambda_{01}\frac{9h_m(Q+L)^2h_m(Q+L')^2}{Z_F^2(Z_m\omega^2+F_{m,k}(L+Q))(Z_m\omega^2+F_{m,k}(L'+Q))(\omega^2+\xi_k(Q)^2)}
\label{eq:flowhd}
\end{align}
\end{widetext}

\subsection{Flow Equation for the Momentum-Dependent Propagators}

The spatial momentum curvature at the minimum of the propagator $A_m,A_d$ can be deduced by solving the flow equation for the momentum dependent propagators $P_m,P_d$.
A straightforward method to obtain the flow of $A_m,A_d$ is a derivative projection. However, we employ a finite difference projection, because it turned out to enhance the stability in our numerical calculations. We chose to evaluate the propagator at a finite distance from the minimum $P=(0,p,0)$, with $p=0.5$. The results are practically independent of the choice of $p$ as long as it is $0\ll p \ll \pi$.
\begin{align}
A_m&=\left(P_m((0,\pi+p,\pi))-P_m((0,\pi,\pi))\right)/p^2\nonumber \\
A_d&=\left(P_d((0,p,0))-P_d((0,0,0))\right)/p^2
\end{align}
\begin{align}
&\partial_k P_m(P)= \tilde \partial_k\sum_Q\frac{1}{4\pi}\nonumber \\
&\Big(  \frac{5\lambda_{20}}{Z_m \omega^2+F_{m,k}(Q)}+\frac{2\lambda_{11}}{Z_d \omega^2+F_{d,k}(Q)}\nonumber \\
&-\frac{4 h_m^2(P)(\omega^2-\xi_k(Q)\xi_k(Q+P))}{Z_F^2(\omega^2+\xi_k(Q)^2)(\omega^2+\xi_k(Q+P)^2)} \Big)
\end{align}
\begin{align}
&\partial_k P_d(P)= \tilde \partial_k\sum_Q\frac{1}{4\pi}\nonumber \\
&\Big(  \frac{3\lambda_{11}}{Z_m \omega^2+F_{m,k}(Q)}+\frac{4\lambda_{02}}{Z_d \omega^2+F_{d,k}(Q)}\nonumber \\
&-\frac{4 h_d^2F_{d,k}(Q+P/2)^2(\omega^2+\xi_k(Q)\xi_k(Q+P))}{Z_F^2(\omega^2+\xi_k(Q)^2)(\omega^2+\xi_k(Q+P)^2)} \Big)
\label{eq:flowmd}
\end{align}

\subsection{Flow Equations for Wave Function Renormalizations}
The flow equations for the wave function renormalizations $Z_m,Z_d,Z_F$ of the bosons and fermions can be extracted from the flow equations of the corresponding propagator. For this purpose we evaluate the flow of the propagators at a finite momentum $P=(\omega_p,0,0)$ from the minimum of the corresponding propagator. We choose $\omega_p=0.5$.
\begin{align}
Z_m&=\left(P_m((\omega_p,\pi,\pi))-P_m((0,\pi,\pi))\right)/\omega_p^2\nonumber \\
Z_d&=\left(P_d((\omega_p,0,0))-P_d((0,0,0))\right)/\omega_p^2\nonumber \\
Z_F&=\left(P_F((\omega_p,0,0))-P_F((0,0,0))\right)/(i\omega_p)
\end{align}
 In the incommensurate case the flow equations are evaluated at the minimum of the magnetic propagator $(0,\pi+\delta_{ic},\pi)$. This definition leads to the corresponding flow equations
\begin{align}
&\partial_k Z_m= \tilde \partial_k\sum_Q\frac{1}{4\pi \omega_p^2}\nonumber \\
&\Big( \frac{4 h_m(\Pi)^2(\omega(\omega+\omega_p)-\xi_k(Q)\xi_k(Q+\Pi))}{Z_F^2(\omega^2+\xi_k(Q)^2)((\omega+\omega_p)^2+\xi_k(Q+\Pi)^2)}\nonumber \\
&-\frac{4 h_m(\Pi)^2(\omega^2-\xi_k(Q)\xi_k(Q+\Pi))}{Z_F^2(\omega^2+\xi_k(Q)^2)(\omega^2+\xi_k(Q+\Pi)^2)}\Big) \ ,
\end{align}
where in the incommensurate case all fermionic kinetic terms $\xi_k(Q+\Pi)$ are shifted to $\xi_k(Q+\Pi+(0,\delta_{ic},0))$.
\begin{align}
&\partial_k Z_d= \tilde \partial_k\sum_Q\frac{1}{4\pi \omega_p^2}\nonumber \\
&\Big( \frac{4 h_d^2F_{d,k}(Q)(\omega(\omega+\omega_p)+\xi_k(Q)^2)}{Z_F^2(\omega^2+\xi_k(Q)^2)((\omega+\omega_p)^2+\xi_k(Q)^2)}\nonumber \\
&-\frac{4 h_d^2F_{d,k}(Q)}{Z_F^2(\omega^2+\xi_k(Q)^2)}\Big)
\end{align}
\begin{align}
&\partial_kZ_F= \tilde \partial_k\sum_Q\frac{1}{8\pi \omega_p}\nonumber \\
&\Big(\frac{6h_m(Q)^2(\omega+\omega_p)}{Z_F(Z_m \omega^2+F_{m,k}(Q))((\omega+\omega_p)^2+\xi_k(Q))}\nonumber \\
&-\frac{6h_m(Q)^2(\omega-\omega_p)}{Z_F(Z_m \omega^2+F_{m,k}(Q))((\omega-\omega_p)^2+\xi_k(Q))}\nonumber \\
&+\frac{4 h_d^2 f(Q/2)^2(\omega+\omega_p)}{Z_F(Z_d \omega^2+F_{d,k}(Q))((\omega+\omega_p)^2+\xi_k(Q)^2)}\nonumber \\
&-\frac{4 h_d^2 f(Q/2)^2(\omega-\omega_p)}{Z_F(Z_d \omega^2+F_{d,k}(Q))((\omega-\omega_p)^2+\xi_k(Q)^2)} \Big)
\end{align}

\section{Regulator}
\label{sec:regulator}

The fermionic propagator diverges at the Fermi surface
\begin{align}
Z_F^2\xi(Q)\xi(Q+\Pi)-\Delta_a=0 \ , \Delta_a=2h_m(\Pi)^2\rho_a \ .
\end{align}
Thus it is imperative to regularize the fermionic propagator at the Fermi surface. In the presence of a nonzero antiferromagnetic condensate the geometry of the Fermi surface changes, which requires the regulator to capture this deformation. Exemplary we demonstrate here how the regulator \eq{eq:regulator} removes the divergence at the Fermi surface. $\xi(\Pi),\xi(\Pi+Q)$ occur interchangeably as a product, thus at fixed $Q$ the Fermi surface is regularized by
\begin{align}
Z_F^2\xi_k(Q)\xi_k(Q+\Pi)-\Delta_a=0 \ ,
\end{align}
where the kinetic terms $Z_F\xi_k(Q)=Z_F\xi(Q)+R_F(\xi(Q))$ and $Z_F\xi_k(Q+\Pi)=Z_F\xi(Q+\Pi)+R_F(\xi(Q+\Pi))$ are regularized by different terms of the regulator, which are without loss of generality
\begin{align}
&R_F(\xi(Q))=Z_F\,\text{sign}(\xi(Q)-\frac{\Delta_a}{Z_F^2\xi_k(Q+\Pi)})\nonumber \\
&\hspace{0.2cm}\times(k-|\xi(Q)-\frac{\Delta_a}{Z_F^2\xi_k(Q+\Pi)}|)\Theta(k-|\xi(Q)-\frac{\Delta_a}{Z_F^2\xi_k(Q+\Pi)}|)\nonumber \\
&R_F(\xi(Q+\Pi))=Z_F\,\text{sign}(\xi(Q+\Pi))\nonumber \\
&\hspace{1cm}\times(k-|\xi(Q+\Pi)|)\Theta(k-|\xi(Q+\Pi)|)  \ . 
\label{eq:reg2}
\end{align}
We now examine what happens if the fermionic kinetic terms get too small and come too close to the Fermi surface. The first regulator in \Eq{eq:reg2} is responsible for introducing a gap at the Fermi surface
\begin{align}
&Z_F^2\xi_k(Q)\xi_k(Q+\Pi)-\Delta_a=\nonumber \\
&Z_F^2\xi_k(Q+\Pi)\,\text{sign}(\xi(Q)-\frac{\Delta_a}{Z_F^2\xi_k(Q+\Pi)})k \ .
\end{align}
The second regulator in \Eq{eq:reg2} acts if the kinetic terms get too close to zero
\begin{align}
&Z_F^2\,\text{sign}(\xi(Q+\Pi))k\,\text{sign}(\xi(Q)-\frac{\Delta_a}{Z_F^2\xi_k(Q+\Pi)})k\nonumber \\
&=Z_F^2\,\text{sign}(Z_F^2\xi(Q+\Pi)\xi(Q)-\Delta_a)k^2 \ .
\end{align}
We have used here that $\text{sign}(\xi(Q+\Pi))=\text{sign}(\xi_k(Q+\Pi))$. An important property is that the Fermi surface is always approached with the correct sign.

\bibliographystyle{apsrev4-1}
\bibliography{BibCollection}


\end{document}